# Novel diamagnetic garnet-type substrate single crystals for ultralow-damping yttrium iron garnet $Y_3Fe_5O_{12}$ films at cryogenic temperatures

C. Guguschev[a,*], C. Dubs[b], R. Blukis[a], O. Surzhenko[b], M. Brützam[a],
R. Koc[a], C. Rhode[a], K. Berger[a], C. Richter[a], C. Berryman[a], R. O. Serha[c,d], A. V. Chumak[c]

[a] Leibniz-Institut für Kristallzüchtung, Max-Born-Str. 2, 12489 Berlin, Germany
[b] INNOVENT e.V. Technologieentwicklung, Prüssingstr. 27B, 07745 Jena, Germany
[c] University of Vienna, Faculty of Physics, Boltzmanngasse 5, 1090 Vienna, Austria
[d] University of Vienna, Vienna Doctoral School in Physics, Boltzmangasse 5, 1090 Vienna, Austria

Abstract

$Y_3Sc_2Ga_3O_{12}$-$Y_3Sc_2Al_3O_{12}$ and $Y_3Sc_2Ga_3O_{12}$-$Y_3Al_5O_{12}$ (YSGAG) solid solution single crystals with diameters up to 30 mm and total lengths up to about 100 mm were grown by the conventional Czochralski technique. Rocking curve measurements on polished sections revealed typical FWHM values of about 22 arcsec, which is indicative of relatively high structural quality for a solid-solution crystal.
The grown substrate crystals are nearly lattice-matched with $Y_3Fe_5O_{12}$ (YIG) to allow epitaxial growth of high-quality thin films. Single crystalline YIG films with thicknesses between 100 nm and 2.9 µm were successfully grown on epi-polished YSGAG substrates using liquid phase epitaxy (LPE). Selected magnetic and microwave properties of the epitaxial films, which still exhibit small lattice misfits to the substrates, were then studied at room temperature. In addition, initial low-temperature investigations confirm that the YIG/YSGAG system is superior to the conventional YIG/GGG ($Gd_3Ga_5O_{12}$) system at temperatures below 10 K, as the ferromagnetic resonance (FMR) linewidth does not increase with decreasing temperature. Therefore, the novel diamagnetic substrates are better suited for microwave applications at low temperature, as excessive damping losses induced by paramagnetic substrates can be avoided. It therefore seems to be a suitable pathway to achieve scalable microwave components for hybrid-integrated quantum systems based on ultralow-damping YIG films that can operate efficiently at millikelvin temperatures.

1. Introduction

Typically, single-crystalline yttrium iron garnet ($Y_3Fe_5O_{12}$; YIG) films with extremely low magnetic damping at room temperature can be grown on commercially available gadolinium gallium garnet ($Gd_3Ga_5O_{12}$; GGG) substrates or other rare-earth substituted gallium garnets [1], provided the lattice misfit allows an epitaxial growth with high crystalline perfection. However, for cryogenic temperature applications, the usually observed increase in microwave damping at low temperatures prevents the use of garnet substrates that exhibit paramagnetic ions in their lattice. In the case of GGG, the paramagnetic moment of the gadolinium ions at low temperatures leads to a stronger increase in magnetization in the substrate compared to that of the YIG film, inducing an opposite inhomogeneous magnetic stray field in the film, which leads to a significant increase in microwave damping [2-4]. In order to avoid such damping contributions at cryogenic temperatures below about 10 K, the complete substitution of the Gd ion by a diamagnetic ion like Y and the partial substitution of Ga by Al and Sc can be a solution to create a completely, or nearly lattice-matched diamagnetic substrate.

Such a material is pivotal for the emerging field of quantum magnonics, which aims to process information using entangled single magnons at millikelvin temperatures [5-9,1]. The availability of diamagnetic substrates enables a transition from substrate-free YIG spheres [10-12] to thin-film platforms, thereby permitting quantum operations with propagating rather than standing – magnons that carry information in space [13-15].

$Y_3Sc_2Ga_3O_{12}$-$Y_3Sc_2Al_3O_{12}$ and $Y_3Al_5O_{12}$-$Y_3Sc_2Ga_3O_{12}$ mixed crystals [16,17] seemed very promising, since mixed crystals have already been demonstrated in the past. However, the crystals reported so far did not have the lattice parameters needed for perfect lattice matching with epitaxial YIG films, since the studies were done in other areas of the respective pseudo-binary systems. Our newly developed "YSGAG" substrate crystals from both systems meet the requirements for a low-loss and lattice-matched substrate that can be applied at cryogenic temperatures. This was confirmed for the first YIG films grown on YSGAG substrates using LPE.

2. Experimental

2.1 Bulk single crystal growth

For the preparation of the starting materials, dried powders of $Sc_2O_3$, $Ga_2O_3$ and $Al_2O_3$ with purities of 99.99% (4N) and $Y_2O_3$ with a purity of 99.999% (5N) were used. Subsequently, cylindrical bars from these powders were made by mixing and then cold isostatic pressing at 0.2 GPa to optimize the crucible filling process. The crystals were grown in a conventional RF-heated Czochralski setup with automatic diameter control. 99.999% pure argon or argon-oxygen mixtures (0.5 to 2.9 % $O_2$) were used as growth atmospheres at atmospheric pressure. Iridium crucibles (inner diameter: 38 and 58 mm) embedded in $ZrO_2$ and $Al_2O_3$ insulation were used. An actively heated iridium afterheater and a lid with an opening were placed on top of the crucible. Crystal growth occurred at pulling rates between 0.4 and 1 mm h$^{-1}$ and at rotation rates between 15 and 23 rpm. Due to the lack of a seed crystal, an iridium rod was used to initiate crystal growth for crystal Y1, which led, after grain selection and competitive growth, to a pulling direction of about 11° off from ⟨110⟩. ⟨111⟩-oriented YAG seed crystals were used for most of the experiments. Whereas the orientation of the seed was maintained for the bulk crystal Y2, Y1.2 contained two individual large grains (the main grain has an orientation of about 22° off from ⟨111⟩). The crystals Y3 to Y13 were single-crystalline. The relevant crystal growth conditions are summarized in Table 1.

Table 1    Crystal growth conditions.

| Crystal | Atmosphere | Growth rate (mm h$^{-1}$) | Rotation rate (rpm) | Used seed | Starting composition |
|---|---|---|---|---|---|
| Y1 | 5N Ar | 0.8 | 20 | Ir rod | $Y_3Sc_2Ga_{1.98}Al_{1.02}O_{12}$ |
| Y1.2 | Ar, 1 vol% $O_2$ | 0.7 | 16 | ⟨111⟩-YAG | $Y_3Sc_2Ga_{1.98}Al_{1.02}O_{12}$* |
| Y2 | Ar, 2.9 vol% $O_2$ | 0.7 | 16 | ⟨111⟩-YAG | $Y_3Sc_{1.7}Ga_{2.55}Al_{0.75}O_{12}$ |
| Y3 | Ar, 0.5 % $O_2$ | 0.6 | 16 | ⟨111⟩-YAG | $Y_3Sc_2Ga_{1.96}Al_{1.04}O_{12}$ |
| Y4 | Ar, 0.5 % $O_2$ | 0.6 | 16 | ⟨111⟩-YAG | $Y_3Sc_{1.7}Ga_{2.55}Al_{0.75}O_{12}$ |
| Y5 | Ar, 0.93 % $O_2$ | 0.6 | 16-23 | ⟨111⟩-YAG | $Y_3Sc_{1.7}Ga_{2.55}Al_{0.75}O_{12}$ |
| Y6 | Ar, 0.5 % $O_2$ | 1 | 15-20 | ⟨111⟩-YAG | $Y_{2.98}Sc_{1.594}Ga_{2.606}Al_{0.82}O_{12}$ |
| Y7 | Ar, 0.28 % $O_2$ | 0.4 | 15-20 | ⟨111⟩-YAG | $Y_{2.982}Sc_{1.718}Ga_{1.9}Al_{1.4}O_{12}$ |
| Y13 | Ar, 0.5 % $O_2$ | 1.0 | 15-20 | ⟨111⟩-YAG | $Y_{2.98}Sc_{1.56366}Ga_{2.54049}Al_{0.919585}O_{12}$ |

* reused melt from run Y1

### 2.2    Powder diffraction

The lattice parameters of the crystal sections close to the seed (Y1 and Y2) and for the beginning and end of the cylindrical parts (Y3 to Y13) were determined by powder X-ray diffraction.

X-ray diffraction measurements on Y1 and Y3-Y13 were performed with a STOE STADI MP diffractometer with a Mo X-ray source fitted with a curved Ge (111) monochromator ($K_{\alpha 1}$ = 0.7093 Å) and a DECTRIS MYTHEN detector in a flat plate transmission geometry. Diffraction patterns were measured over a range of 2–50 degrees 2Θ (Q = 0–8 Å$^{-1}$) for a duration of 0.25 h per data point. Samples Y1 and Y3 to Y7 were prepared by collecting the powder created by drilling selected sub-surface locations of single crystal boules with a diamond drill mortar. Samples for Y13 were prepared by grinding pieces of single crystal wafers in an agate mortar.

X-ray diffraction measurement for Y2 was performed with a STOE STADI P diffractometer with a Cu X-ray source fitted with a curved Ge (111) monochromator ($K_{\alpha 1}$ = 1.5406 Å) and a DECTRIS MYTHEN2R detector in a flat plate transmission geometry. The diffraction pattern was measured over a range of 0–116 degrees 2Θ (Q = 0–6.92 Å$^{-1}$) for a duration of 0.5 h per data point. The sample was prepared by grinding a piece of a single crystal wafer in an agate mortar. Powdered Si (1.73 wt%) was added as an internal lattice parameter standard.

The samples were analysed by Rietveld refinements performed using GSAS-II [16]. During the refinement unit cell, microstrain (grain size was assumed to be infinite), site occupancies and isotropic thermal displacements were optimized. The background was approximated using Chebyshev polynomials, and the instrument function was calibrated by empirically fitting the Caglioti function with included asymmetry to an NIST Si measured in the same geometry.

### 2.3 Chemical and microstructural analyses

Micro X-ray fluorescence (µ-XRF) elemental mapping and energy dispersive Laue mapping (EDLM) were carried out under low vacuum conditions (20 mbar) using a Bruker M4 TORNADO spectrometer to: (1) investigate the chemical homogeneity of polished (111)-oriented sections prepared from the grown crystals and (2) assess the microstructure.

EDLM has proven capabilities to detect low- and high-angle grain boundaries, twins, striations and dislocations in crystalline materials [19-22]. The measurement system was equipped with a rhodium X-ray source operated at 50 kV and between 200 and 600 µA. Polycapillary X-ray optics were used to focus the non-polarized white radiation at the surface of the sample, resulting in a spatial resolution of about 20 µm. The white radiation from the excitation source interacts with the crystals and due to the instrument geometry, it is possible to detect Bragg reflections. The principles of the XRD surface mapping technique, the measurement procedure and the measurement setup are described elsewhere [19].

Analysis of the sample areas for Y2 were done at a step size between spot measurements of 10 µm. The surface area was scanned "on the fly" by moving the sample stage continuously. The measurement time per point was set between 18 and 27 ms and two silicon drift detectors (SDD) detectors were used.

For determination of the chemical homogeneity of the cross section from edge to edge and for elemental mappings, the µ-XRF spectrometer was calibrated using ceramic standard reference samples. Point measurements were conducted with a measurement time of 100 s per point. For the determination of the average chemical composition Y1 samples were scanned with 20 ms per point using one detector, 60 µm step size and 3 cycles.

Further chemical compositions were determined for the crystals Y3 and Y13 by inductively coupled plasma optical emission spectroscopy (ICP-OES) using a ThermoFisher iCAP 7400 instrument. The powders (10-20 mg) probed from the beginning and the end of the cylindrical sections of the grown crystals were dissolved in mixtures consisting of 2 mL H$_3$PO$_4$, 1 mL HNO$_3$ and 1 mL H$_2$O$_2$ in polytetrafluoroethylene (PTFE) lined autoclaves using a microwave assisted (240 °C, 30 min) digestion process. The obtained mixtures were then diluted to 50 ml total volume for the measurements. The ICP-OES was calibrated using ROTI®Star ICP-OES mono-elemental standards (Carl Roth GmbH) (monitored lines: Y 371.030 nm; Al 394.401 nm; Ga 294.364 nm; Sc 335.373 nm). Each sample was measured 3 times and a quality check was performed after the samples were measured.

### 2.4 Rocking curve imaging

Rocking curve imaging (RCI) was used to obtain spatially resolved maps of lattice parameters throughout the wafer surface. We used a Rigaku SmartLab high-resolution diffractometer equipped with a HyPix-3000 2D pixel detector to perform RCI with a spatial resolution of ~0.2 mm. The diffractometer uses a rotating anode and CuK$_{\alpha 1}$ radiation ($\lambda$ = 1.5405929(5) Å) selected by a 2-bounce Germanium 400 monochromator. We recorded RCI data for several Bragg reflections and azimuthal positions to disentangle the quantitative information about lattice tilt and lattice parameters. The spatial information was obtained using the HyPix-3000 detector at a distance of 200 mm. The field of view was approx. 8 mm horizontal x 1 mm vertical. In the (vertical) scattering plane, the spatial resolution is determined by the pixel size of the detector (100 µm) divided by the sine of the exit angle which depends on the selected Bragg reflection. Perpendicular to the scattering plane, the spatial resolution is defined by a Soller slit. The angular acceptance was 0.114°, leading to a spatial resolution of approx. 400 µm. To map the entire sample, it was scanned through the beam in the lateral *x-y* directions, taking subsequent images with deliberate overlap. The angular runout of the linear stages has previously been characterized by doing similar measurements on a single crystalline silicon block. The raster scan was repeated for a range of rocking angles covering the complete rocking curve for all parts of the crystal.

The whole procedure was repeated for three different azimuthal positions $\varphi$ (rotation about the lattice-plane normal) of 0°, 90°, 180° for the symmetric 444 reflection. For each point of the sample and each azimuthal position, we determined the peak positions of the rocking scans as well as the integrated intensity and the peak FWHM. The true Bragg angle – and hence the *c* lattice parameter – can be determined by the average of the peak positions for $\varphi$ = 0° and $\varphi$ = 180°. Accordingly, the difference equates to twice the local lattice tilt (pitch) of the normal of the lattice planes. Finally, the difference of the peak positions at $\varphi$ = 90° from the true Bragg angle yields the lattice tilt in the orthogonal direction (roll). In addition, RCI maps were obtained for the asymmetric reflections 264, 624, 336 and 552.

The out-of-plane lattice parameter d$_{111}$ is determined by the measurements of the 444 reflection. Using the assumption of orthorhombic symmetry (no change of lattice angles), the in-plane lattice parameters $d_{1\bar{1}0}$ and $d_{11\bar{2}}$ have been determined by additional measurements of the 264, 624, 336 and 552 reflections, respectively. The results are shown in Fig. 1.

$$d_{1\bar{1}0} = 2\left[\left(\frac{1}{d_{264}}\right)^2 - \left(\frac{1}{d_{444}}\right)^2\right]^{-1/2}$$

$$d_{\bar{1}\bar{1}2} = \left[\left(\frac{1}{d_{336}}\right)^2 - \left(\frac{1}{d_{444}}\right)^2\right]^{-1/2}$$

$$d_{111} = 4d_{444}$$

## 2.5 Thermal analysis methods

The melting points of samples belonging to the crystals Y1 and Y2 were determined with a NETZSCH STA 429 CD device using the high-temperature differential thermal analysis (DTA) method and thermogravimetry (TG). Each sample was placed in a Wolfram crucible and the measurements were conducted in a static He atmosphere with a heating rate of 20 K min$^{-1}$.

## 2.6 Thermal conductivity

Thermal diffusivity was measured with a NETZSCH Laser Flash analysis (LFA) 427 device. As the samples were highly IR-transparent, they were coated on both surfaces to improve IR absorbance and emittance. Pt nanoparticles were used as a coating and the samples were measured in Ar/O$_2$ atmosphere to prevent reduction of Ga(III). Thermal diffusivity was calculated from the raw measurement data of detector voltage versus time by fitting the voltage profile using Mehling's model [18].

Heat capacity was measured with a NETZSCH STA 449 F3 Jupiter® device. The measurement of heat capacity is required to calculate thermal conductivity (Equation 1) and it grants the possibility to detect various phase transitions or diffusion processes happening in the sample at high temperature if present. Measurements were performed in Pt crucibles under Ar/O$_2$ gas flow with a heating rate of 20 K min$^{-1}$ to maximize instrument sensitivity.

$$\kappa(T) = \alpha(T) \times \rho \times C_p(T), \text{ Equation 1}$$

where $\kappa(T)$ is thermal conductivity, $\alpha$ is thermal diffusivity, $\rho$ is density (assumed to be constant) and $C_p(T)$ is heat capacity.

## 2.7 Heteroepitaxial YIG film growth

Several micrometer-thick as well as submicrometer-thin YIG films were deposited on (111) YSGAG substrates by LPE, using PbO-B$_2$O$_3$-based high-temperature solutions in the temperature range between 790°C and 840°C at ambient pressure and the isothermal dipping method (see e.g. [23, 24]). For this purpose, unformatted, one-sided chemo-mechanically polished (CMP) (111) YSGAG substrates prepared from crystal Y1 were used to realize epitaxial growth of YIG films by means of the vertical LPE dipping technique in a vertical, resistively heated three-zone furnace. The deposition was realized by vertical fixation of the substrates with a Pt wire through a hole in the substrate and dipping of the sample in the supercooled solution after a 5-min temperature equilibration phase above the surface of the solution.

Epitaxial YIG films were achieved during dipping times between 1 to 20 minutes with growth rates of about 0.1 – 0.2 µm/min. After deposition, the samples were slowly pulled out of the solution to avoid dynamic wetting of melt residues on the film surfaces, and then quickly pulled out of the furnace to allow them to rapidly cool to room temperature. Subsequently, the sample holder with the sample was stored in a diluted, hot nitric-acetic-acid solution to remove the rest of the solidified solution residues. The film thicknesses were determined using a prism coupler technique with a Metricon Model 2010/M at laser wavelengths of 633 nm and 1550 nm and a measuring spot smaller than 1 mm². Finally, the backside of the sample was removed by mechanical polishing and the samples were cut into chips of different sizes using a diamond wire saw.

### 2.8 X-ray diffraction and magnetic measurements on YIG LPE films

XRD measurements were performed to determine the lattice misfit between the YSGAG substrates and the epitaxial YIG films. Using a Malvern Panalytical PW 3149/63 hybrid monochromator equipped with a (220) Ge-crystal for parallel-beam geometry and transmitting only $K_{a1}$ radiation, θ-2θ X-ray patterns were obtained around the (888) garnet substrate reflections. The detector was a PIXCEL3D 2D-detector with a 0.04 rad soller and a 7.5 mm fixed anti-scatter slit. Using the MalvernPanalytical Highscore Plus V.4.9 program, the measured peaks were fitted by applying the Pearson VII form function to determine the lattice misfit between the film and the substrate.

The vibrating sample magnetometer (VSM, MicroSense LLC, EZ-9) was used to measure the magnetic moments of the YIG/GGG samples magnetized along the YIG film surface. The external magnetic field $H$ was controlled with an error of ≤0.01 Oe. To determine the volume magnetization $M$ of the YIG films, the raw VSM signal was corrected for background contributions (due to the sample holder and the substrate) and normalized to the YIG volume.

The absorption spectra of the ferromagnetic resonance (FMR) at room temperature were carried out in an in-plane setup. The frequency-swept measurements were recorded with a Rohde & Schwarz ZVA 67 vector network analyzer connected to a broadband stripline. The sample was mounted face-down on the stripline, and the transmitted signals $S_{21}$ and $S_{12}$ were recorded using a source power of -10 dBm (0.1 mW). The microwave frequency was swept across the resonance frequency $f_{res}$, while the in-plane magnetic field $H$ remained constant. Each recorded frequency spectrum was fitted with a Lorentz function and allowed us to determine the resonance frequency $f_{res}$ and the frequency linewidth $\Delta f_{FWHM}$ according to the applied field $H = H_{res}$.

The temperature-dependent absorption spectra were performed within a Physical Property Measurement System (PPMS) at temperatures ranging from 2 K to 300 K and up to 40 GHz. The sample was mounted on a frequency-broadband stripline perpendicular to the YIG stripe and placed in a homogeneous magnetic field of up to 1.3 T, generated by superconducting coils. The maximum applied microwave power in the PPMS was -5 dBm. For temperatures below 2 K, measurements were conducted in a dilution refrigerator, achieving base temperatures around 10 mK. At 20 mK, the refrigerator provides a cooling power of 14 µW, which is sufficient to maintain thermal equilibrium during FMR spectroscopy measurements with an applied power of -25 dBm. To obtain the FMR resonance frequency and full linewidth at half maximum (FWHM) of the resonance rf-power absorption peak, the resonance curve is fitted using a Lorentzian model.

3. Results and discussion

3.1 Crystal growth

Garnet-type $Y_3Sc_2Ga_3O_{12}$-$Y_3Sc_2Al_3O_{12}$ and $Y_3Sc_2Ga_3O_{12}$-$Y_3Al_5O_{12}$ (YSGAG) solid-solution crystals with diameters up to 30 mm and total lengths up to 100 mm were grown by the Czochralski technique. The first grown crystals Y1 and Y2 and the polished substrates were found to be transparent (water-clear) (Fig. 1a,c and Fig. 2), except the tail part of crystal Y1.2 (Fig. 1c) that was translucent, which indicates scattering centers and potential inclusions. The dark spots at the surfaces of the crystals Y1.2 and Y2 are iridium-containing deposits grown from the vapor phase, which are more intense for crystal Y2 since it was grown at a higher oxygen partial pressure. Contamination by $Nd^{3+}$ was found in crystals Y1 and Y1.2, causing a slight coloration for the latter crystal. This contamination could originate from a pre-used crucible for crystal growth. The second and third generation of crystals (Y3 to Y7, Fig. 3-4) were of high quality without significant amounts of impurities or scattering centers. New crucibles were used for these experiments. Some specifications of the grown crystals are summarized in Table 2.

The surfaces of the crystals were partially non-transparent and irregular as a result of thermal etching, by the interaction with Ga-related gaseous species ($Ga_2O$, $GaO$ and $Ga$), and deposition of Ir particles/crystals originating from oxidation of the crucible material.

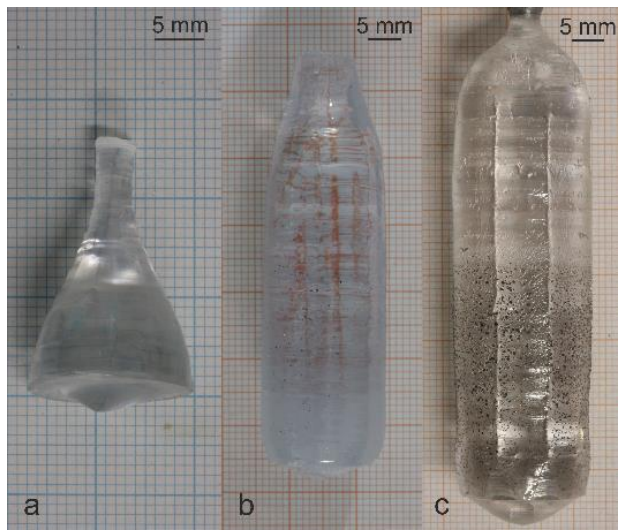

Figure 1: First series of Czochralski-grown $Y_3Sc_2Ga_3O_{12}$-$Y_3Sc_2Al_3O_{12}$ and $Y_3Sc_2Ga_3O_{12}$-$Y_3Al_5O_{12}$ (YSGAG) solid solution single crystals (a: crystal Y1, b: crystal Y1.2, c: crystal Y2).

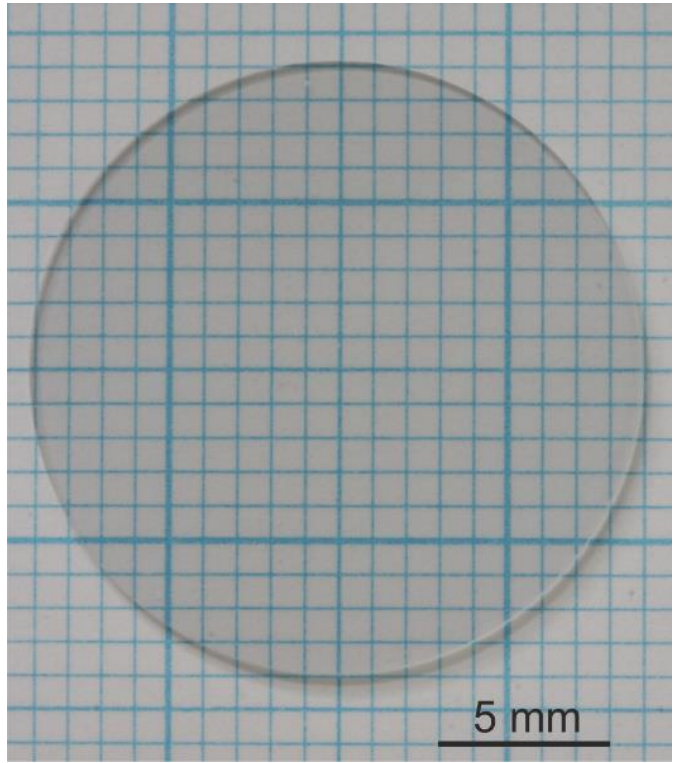

Figure 2: Chemo-mechanical polished (111) substrate prepared from crystal Y2 with a diameter of 18 mm.

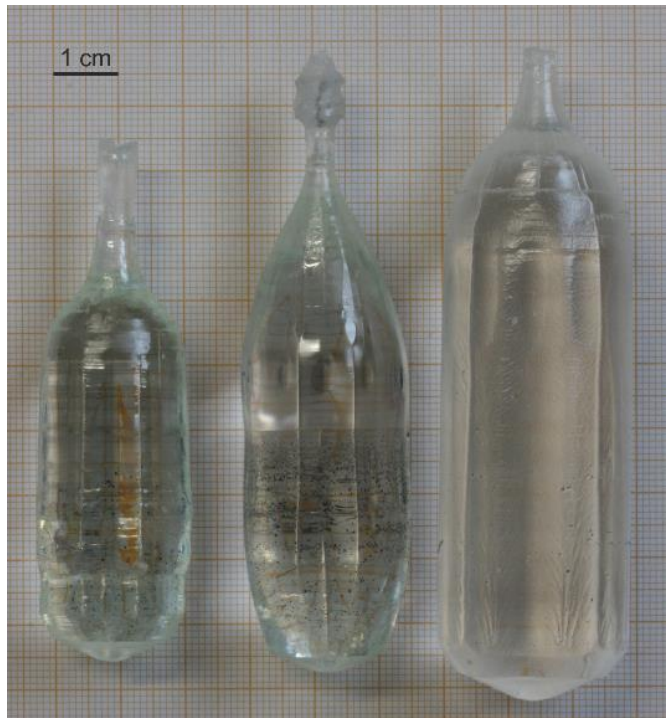

Figure 3: Second series of Czochralski-grown $Y_3Sc_2Ga_3O_{12}$-$Y_3Sc_2Al_3O_{12}$ and $Y_3Sc_2Ga_3O_{12}$-$Y_3Al_5O_{12}$ (YSGAG) solid solution single crystals (left: Y3, mid: Y4, right: Y5).

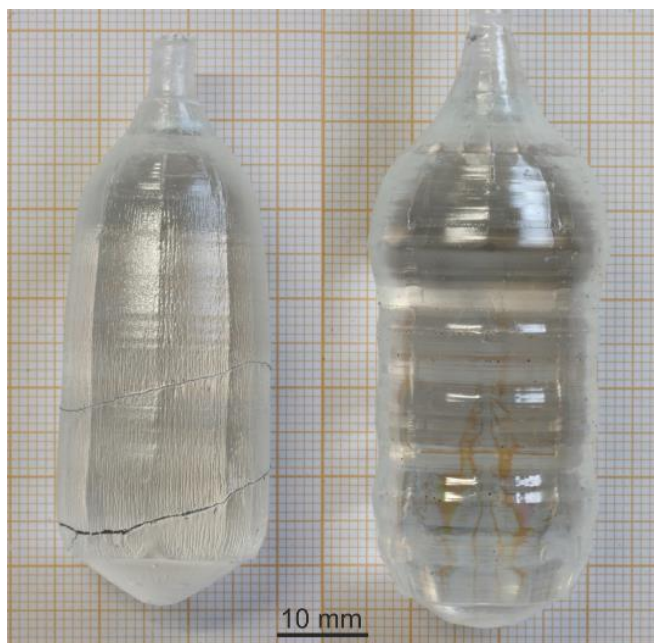

Figure 4: Third series of Czochralski-grown YSGAG solid solution single crystals (left: Y6, right: Y7) with adjusted compositions.

Table 2: Specifications of the grown crystals

| Crystal | Max. crystal diameter (mm) | Crystal length (mm) | Cylinder length (mm) | Solidified fraction |
|---|---|---|---|---|
| Y1 | 17 | 25 | - | < 0.1 |
| Y1.2 | 18 | 62 | ~40 | ~0.4 |
| Y2 | 20 | 77 | ~60 | ~0.6 |
| Y3 | 22 | 80 | ~40 | ~0.6 |
| Y4 | 25 | 95 | ~40 | ~0.8 |
| Y5 | 30 | 100 | 65 | ~0.4 |
| Y6 | 23 | 63 | ~35 | ~0.5 |
| Y7 | 26 | 78 | ~35 | ~0.7 |
| Y13 | 24 | 55 | ~35 | ~0.6 |

3.2 Chemical composition of selected crystals

Chemical data relying on the reference calibrated micro X-ray fluorescence (µ-XRF) results are summarized in Table 3 for the crystals Y1 and Y2. The elements Nd and La (only present as impurities in the Y1 and Y2 crystals) were neglected. µ-XRF shows high precision, but less accurate results for these very complex systems, primary caused by Al fluorescence absorption issues in these mixed crystals.

More reliable chemical compositions were measured by inductively coupled plasma optical emission spectroscopy (ICP-OES) for the crystals Y3 and Y13 (see Table 2), which are good representatives for optimized crystal compositions (for nearly lattice matched substrates for YIG) within the $Y_3Sc_2Ga_3O_{12}$-$Y_3Sc_2Al_3O_{12}$ and $Y_3Sc_2Ga_3O_{12}$-$Y_3Al_5O_{12}$ systems, respectively. Thereby A denotes the beginning of cylindrical part and E denotes the end of the cylindrical part.

Table 3: Chemical composition of selected crystals obtained by µ-XRF and ICP-OES. A denotes beginning of cylindrical part of the crystal and E denotes end of cylindrical part of the crystal. The shown errors are sigma values.

|  | Y (at.%) | Sc (at.%) | Ga (at.%) | Al (at.%) |
|---|---|---|---|---|
| Y1 starting material | 15.00 | 10.0 | 9.90 | 5.10 |
| Y1 crystal (shoulder), µ-XRF | 14.91 | 8.59 | 9.50 | 7.00 |
| Y2 starting material | 15.00 | 8.50 | 12.75 | 3.75 |
| Y2 crystal (cylindrical part), µ-XRF | 14.78 | 7.90 | 13.06 | 4.26 |
| Y3 A, ICP-OES | 15.07 ± 0.04 | 8.98 ± 0.18 | 9.593 ± 0.015 | 6.356 ± 0.004 |
| Y3 E, ICP-OES | 15.06 ± 0.03 | 9.33 ± 0.09 | 9.49 ± 0.03 | 6.123 ± 0.022 |
| Y13 A, ICP-OES | 15.14 ± 0.15 | 7.83 ± 0.15 | 11.89 ± 0.10 | 5.15 ± 0.04 |
| Y13 E, ICP-OES | 15.04 ± 0.06 | 8.07 ± 0.16 | 11.96 ± 0.05 | 4.930 ± 0.023 |

For a polished substrate of crystal Y2 with a diameter of 18 mm, we found a high chemical homogeneity, as shown in Figure 5a-b. Fig. 5a shows the concentration profiles of Yttrium (Y), Gallium (Ga), Scandium (Sc) and Aluminum (Al) in descending order.

A ten-fold determination of the same point at the sample revealed sigma values of 0.049, 0.022, 0.028, 0.049 for Y, Sc, Ga, Al, respectively. These results are comparable to the sigma values obtained by the linescan across the surface of the substrate (0.027, 0.017, 0.022, 0.057 were determined for Y, Sc, Ga, Al, respectively), i.e. the chemical homogeneity of this substrate is comparatively high for a solid solution crystal.

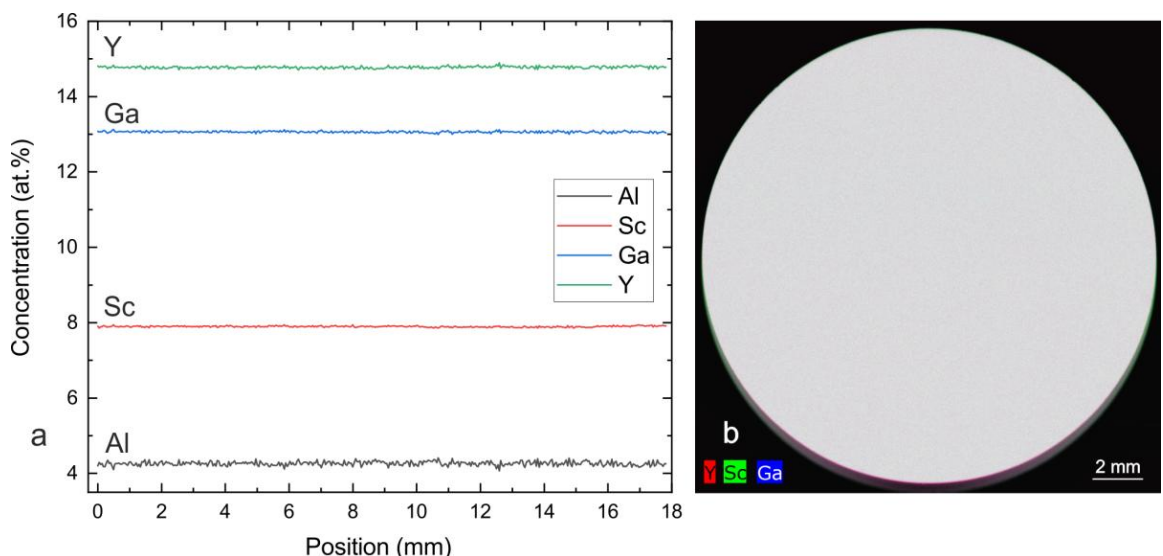

Figure 5: (a) Chemical composition measured on a (111)-oriented polished disc of crystal Y2 with a diameter of 18 mm from edge to edge through the center. (b) Superimposed color-coded intensity plot of Y, Sc and Ga across the entire area of the cross-sectional piece from the cylindrical part of the same crystal.

In relation to the segregation profiles obtained along the pulling directions it was found that for typical Czochralski yields up to solidified fractions of about 0.5 the segregation is of minor extent for the crystals Y4, Y5 and Y6, which are based or modified from the $Y_3Al_5O_{12}$-$Y_3Sc_2Ga_3O_{12}$ system. The $Y_3Sc_2Ga_3O_{12}$-$Y_3Sc_2Al_3O_{12}$ mixed crystal Y3 and the crystal Y7 grown from an adjusted composition revealed a lower fraction between 0.3 and 0.35, respectively.

### 3.3 Structural data

The powder XRD measurements revealed that all measured samples from all generations consist of a pure garnet phase (Figures 6). The lattice parameters were determined to be 12.3874 Å and 12.4016 Å for the samples originating from the Y1 and Y2 crystals, respectively. The results of additional measurements for the second and third generation of crystals is shown in Table 4. The determined lattice parameters lie between 12.348 and 12.4136 Å. Here, the lattice parameter shifts within the cylindrical sections for the start (A) and end sections (E)) were determined. According to this data and the chemical information (see above under paragraph 3.2) it is obvious that a segregation is significantly lower for the crystals grown in the $Y_3Al_5O_{12}$-$Y_3Sc_2Ga_3O_{12}$ system compared to the $Y_3Sc_2Ga_3O_{12}$-$Y_3Sc_2Al_3O_{12}$ mixed crystals.

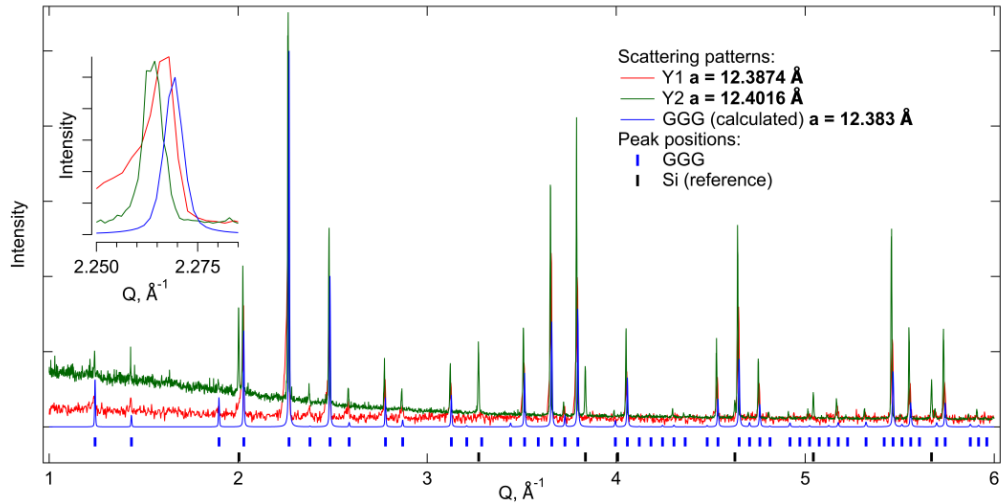

Figure 6: X-ray powder diffractograms for Y1 and Y2 compared to GGG.

Table 4: Obtained lattice parameters of the crystals of the second and third series (A: start of cylinder, E: end of cylinder)

| Crystal | Lattice parameter (Å) | Error (Å) |
|---|---|---|
| Y1 | 12.3874 | +0.009/-0.002 |
| Y2 | 12.4016 | +0.009/-0.002 |
| Y3 A | 12.3777 | +0.009/-0.002 |
| Y3 E | 12.3944 | +0.009/-0.002 |
| Y4 A | 12.3925 | +0.009/-0.002 |
| Y4 E | 12.4136 | +0.009/-0.002 |
| Y5 A | 12.3856 | +0.009/-0.002 |
| Y5 E | 12.3992 | +0.009/-0.002 |
| Y6 A | 12.3795 | +0.009/-0.002 |
| Y6 E | 12.3895 | +0.009/-0.002 |
| Y7 A | 12.3480 | +0.009/-0.002 |
| Y7 E | 12.3651 | +0.009/-0.002 |
| Y13 A | 12.3774 | +0.009/-0.002 |
| Y13 E | 12.3836 | +0.009/-0.002 |

### 3.4 Structural quality and homogeneity

By using Energy-dispersive Laue mapping for several polished wafers at several positions within the first half of the boule Y2 (see Fig. 7), striations and fingerprints of the appearing central facets at the growth interface can be visualized. It is obvious that the geometric dimensions of the facets in the core part are not constant, but change dynamically in the course of crystal growth.

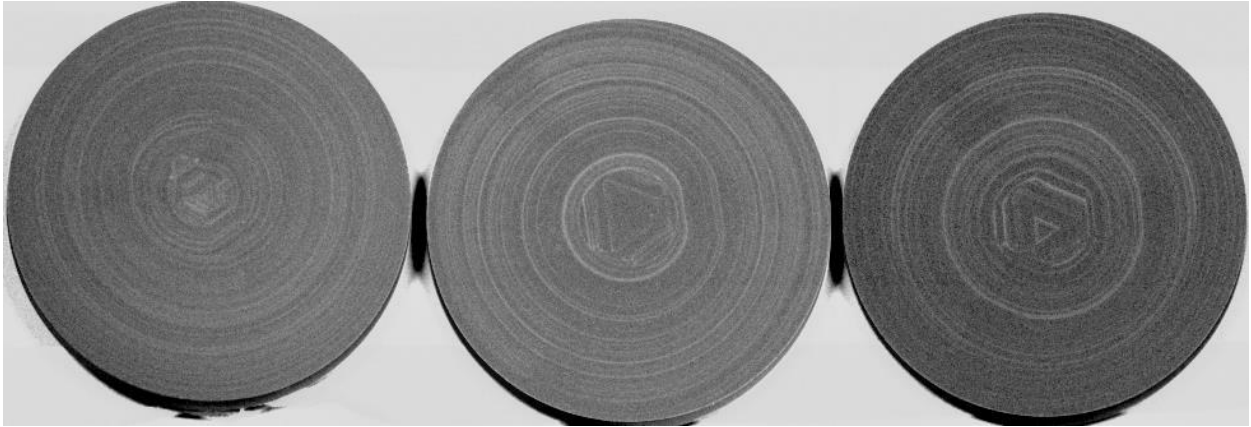

Figure 7: Energy-dispersive Laue mapping results for several polished discs from the first half of the cylindrical part of crystal Y2 (superimposed greyscale-coded 2D intensity plot of selected Bragg reflections).

Similar features can be also seen by the RCI plots in Figs. 8 and 9. The fine variations of the d-values can be resolved and quantified by using this technique. Furthermore, the rocking curves of the 444 reflection have been analyzed with respect to their peak width, resulting in the map of the "full width at half maximum" in Fig. 9. An average FWHM value of about 22 arcsec (median: about 20 arcsec) was determined, which indicates a relatively high structural quality for a solid-solution crystal.

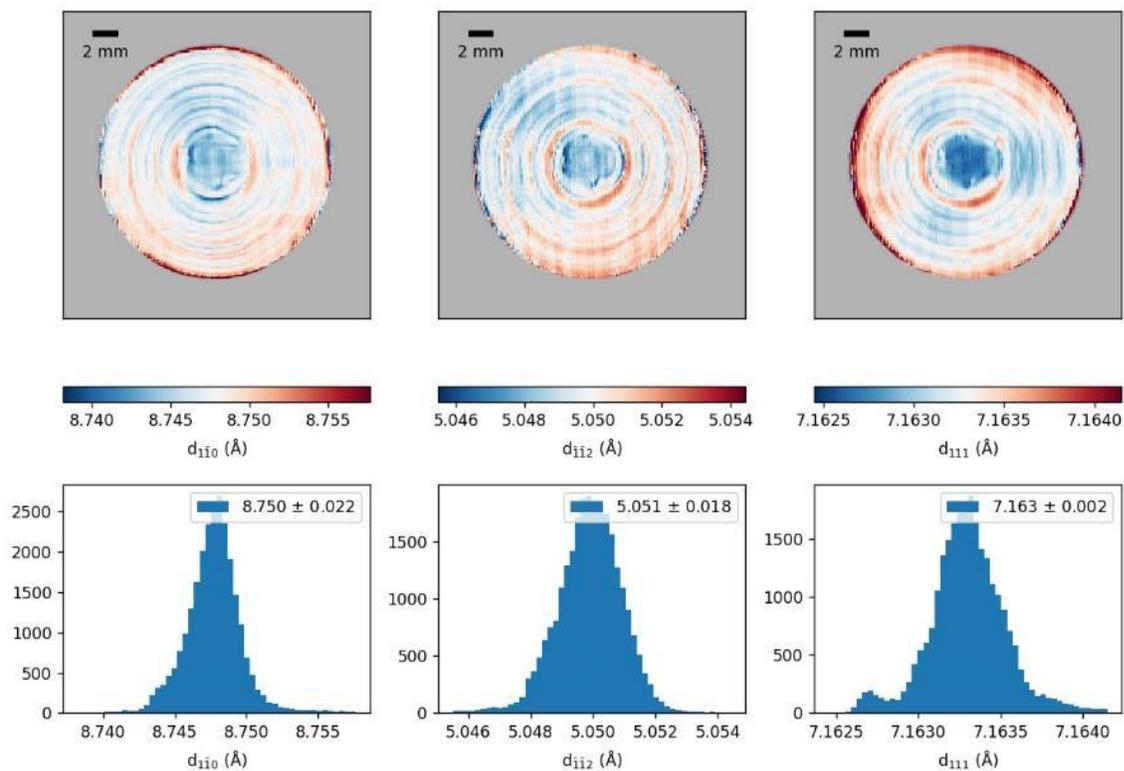

Figure 8: 2D distributions and histograms the obtained d-values for the CMP substrate from crystal Y2.

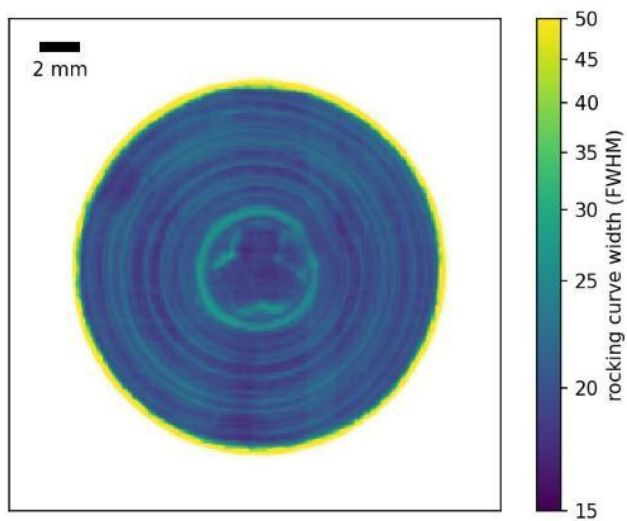

Figure 9: 2D distribution of the FWHM of the 444 reflection for the CMP substrate from crystal Y2.

### 3.5 Melting behavior

The melting points of the samples originating from the crystals Y1 and Y2 were observed to be 1872 ± 10 °C and 1866 ± 10 °C respectively (Figure 10). In both cases the melting was accompanied by a significant mass loss, which was much stronger for Y2 since the gallium content was much higher. This is to be expected as upon melting Ga(III) in the melt gets reduced to the volatile $Ga_2O$ in the He atmosphere.

The temperature was calibrated based on an internal alumina reference melting point that was linearly offset to coincide with 2054 °C for every heating step.

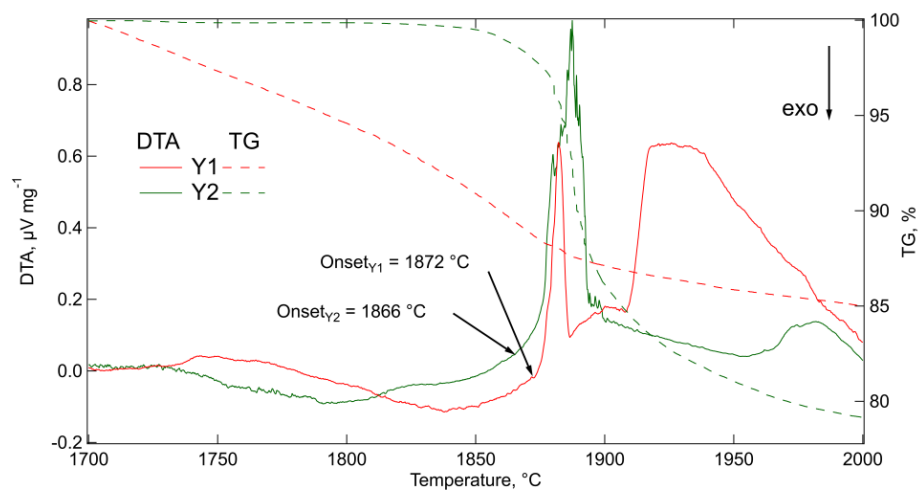

Figure 10. DTA and TG curves of the Y1 and Y2 samples as measured in an inert atmosphere.

### 3.6 Thermal conductivity

Thermal conductivity of both samples shows a decrease with temperature from 6 Wm$^{-1}$K$^{-1}$ (Y1) or 3.9 Wm$^{-1}$K$^{-1}$ (Y2) at room temperature to about 2.8 Wm$^{-1}$K$^{-1}$ and 1.5 Wm$^{-1}$K$^{-1}$ at 1450 °C for Y1 and Y2 respectively (Fig. 11). This is in accordance with an overall expectation of decreasing thermal conductivity at high temperatures due to phonon scattering. Room temperature thermal conductivity is 2-3 times lower than that of YAG (12 Wm$^{-1}$K$^{-1}$) [25]. This is likely due to the solid solution nature of Y-Sc-Ga-Al garnets as it contains different atoms of different masses on each of the cation lattice sites. The origin of the significant difference between the two samples is as of now unclear. However, it is likely due to different occupancies of B and C sites, though it remains active area of work. The same trends as observed in thermal conductivity are also observed in thermal diffusivity, indicating the differences between samples are not caused by incorrect assumptions about densities or heat capacities. No significant features were observed in heat capacity with the peak-like changes and peak at 1200 °C arising due to sample holder artefacts.

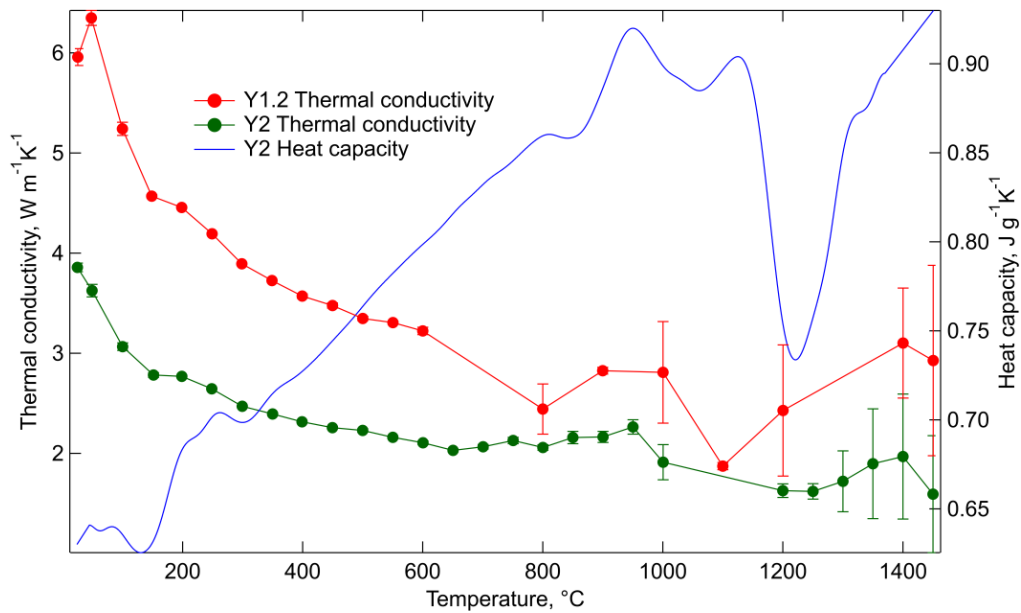

Figure 11. Thermal conductivities (red and green) of the crystals Y1.2 and Y2 and heat capacity (blue) of crystal Y2.

### 3.7 Epitaxial film growth

Liquid phase epitaxy is known to be able to grow particularly perfect single crystalline films with thicknesses in the micrometer range up to hundreds of micrometers. An ideal model system for oxides is the epitaxial growth of magnetic iron garnets on rare earth gallium garnet substrates [14]. Recently, it has been shown that this deposition technique can also be applied to grow high-qualitative, nm-thin YIG films with thicknesses down to ten nanometers [27,28]. These films, which are usually fabricated by the horizontal dipping technique using substrate rotation, are characterized by an exceptionally homogenous film thickness and topographical and microstructural perfection. However, this requires sufficiently large, round substrates. These must be fixed horizontally in substrate holders to enable optimum epitaxial growth and spinning off of the melt residues after leaving the high-temperature solution. However, for unformatted and small substrate samples, the vertical dipping technique is very effective for quickly growing initial films and characterizing their film properties even before substrates with larger diameters and higher structural perfection can be used. Therefore, we used this technique for the very first one-sided epi-polished YSGAG (Y1) substrates available. Due to the vertical dipping technique used, film thickness gradients were observed for all samples e.g. about 10% for micrometer-thick samples and up to 60% for 100 nm to 200 nm thin samples. In order to reduce the effects of the inhomogeneity of the film thickness on the magnetic and microwave properties, square chips with edge lengths of 4 mm or 5 mm were prepared, of which selected specimens are shown in Figure 12.

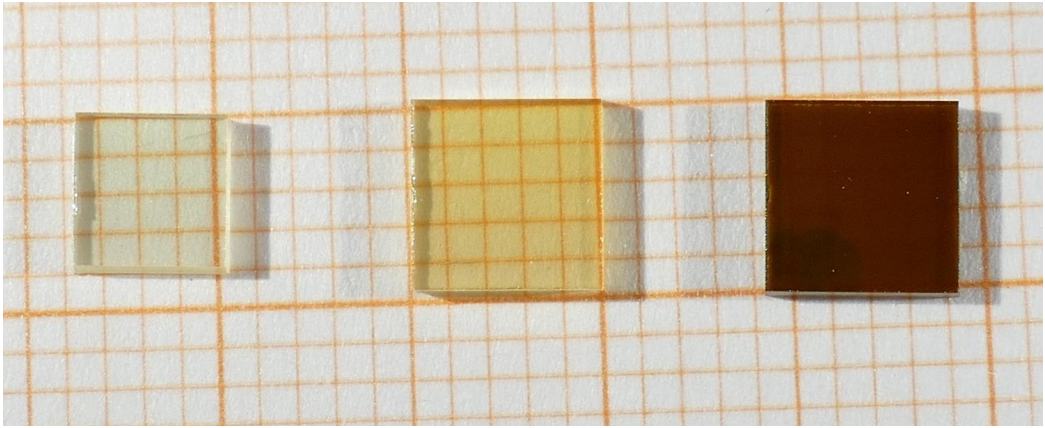

Figure 12: Selected samples of YIG films grown on one-sided epi-polished (111) YSGAG substrates by liquid phase epitaxy using the vertical dipping technique. The edge lengths of the samples are between 4 to 5 mm and the averaged film thicknesses from left to the right are: ~100 nm, ~160 nm and ~2.9 µm.

### 3.8  Lattice misfit

In the case of micrometer-thick films, XRD measurements revealed that both the YSGAG substrate and the YIG film exhibit Bragg reflections with high peak intensities, whereby an angular separation $2\theta$ of ~ 0.40° was determined, resulting in a lattice misfit of $\Delta a_\perp = a_{film} - a_{substrate}$ = -0.0024 nm or a relative misfit $\Delta a_\perp/a = (a_{film} - a_{substrate})/a_{film}$ = of 0.2 % (see Fig. 13a). The same values were obtained for the nanometer-thin films down to a thickness of 100 nm, whereby the peak intensities of the films were significantly lower (see Fig. 13b). While both films were crack-free after deposition, the 2.9 µm film developed cracks after storing. This is an indication that the strain induced by the lattice misfit is too high and that the critical film thickness for crack-free films has been exceeded. For YSGAG substrates prepared from crystal Y1, the substrate lattice parameter is therefore too large for films several micrometers thick. To achieve films several micrometers thick, YSGAG substrates with smaller lattice parameters such as Y3A, Y5A, Y6A and Y13 can be used, as their lattice parameters are close to the usual lattice parameter of the GGG substrate of 1.2383 nm or closer to the bulk lattice parameter of nominal pure YIG of 1.2376 nm. However, submicrometer-thin films should be realizable on all other grown substrate crystals, because their relative lattice misfits are lower than the value of ~0.4 % reported by Guo et al. [29] for 30 nm thin films on commercially available diamagnetic $Y_3Sc_2Ga_3O_{12}$ (YSGG) substrates ($a_{bulk}$ = 1.2426 nm). In the case of the 30 nm-thin, Y-rich, off-stoichiometrically substituted $Y_3(Y_xFe_{5-x})O_{12}$ films reported by Legrand et al. [30], an attempt was made to adapt the lattice parameter of the films to the substituted substrates used, in particular to the diamagnetic YSGG ($a_{measured}$ = 1.2460 nm). Both approaches are very promising for the deposition of high-quality, nanometer-thin iron garnet films on diamagnetic garnet substrates. However, in contrast to Guo et al. [29], the new YSGAG substrates can be perfectly matched to YIG lattice, and in contrast to Legrand et al. [30], the LPE technique has the additional potential to grow not only nanometer-thin but also micrometer-thick films without degradation of crystalline perfection over micrometer thicknesses, which are required in practice for planar microwave devices.

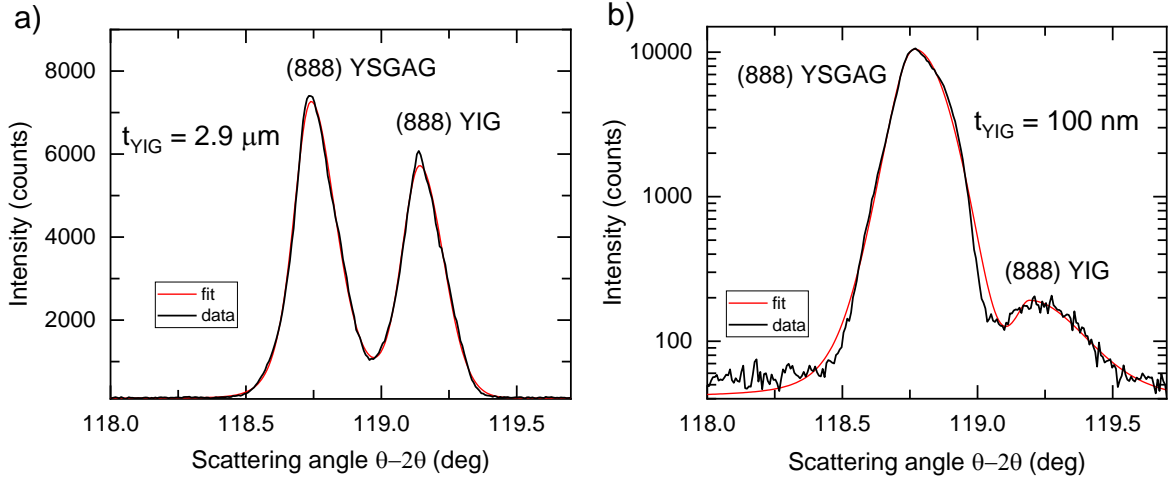

Figure 13: $\theta$-$2\theta$ scans around the 888 Brag reflections using CuK$_{\alpha 1}$ radiation (a) of a 2.9 µm YIG film on a YSGAG substrate and (b) of a about 100 nm YIG film which show the still existing lattice misfit of the epitaxial systems (black curve: measurement, red curve: fit curve).

### 3.9 Magnetic properties

Vibrating sample magnetometry was carried out at room temperature to characterize the magnetic properties of selected samples. For submicrometer-thin YIG/YSGAG films, the measurements reveal hysteresis curves with small coercivities of about $H_c$ = 0.3 Oe (Figure 14a), which are comparable to in-plane magnetized YIG films on commercially available GGG substrates and which can be grown, for instance, by LPE [27,28] or for nanometer-thin films ≤100 nm by PLD [31] or rf-sputtering [32]. For micrometer-thick LPE films on YSGAG substrates, the shape of the hysteresis curve is also comparable to micrometer-thick YIG/GGG samples, which exhibit an abrupt domain wall displacement at small bias fields (hysteresis) and then show a continuous domain rotation at higher fields, resulting in saturation fields $M_s \sim (1650 \pm 150)$ G in the case of the 2.9 µm YIG/YSGAG sample investigated (Figure 14b).

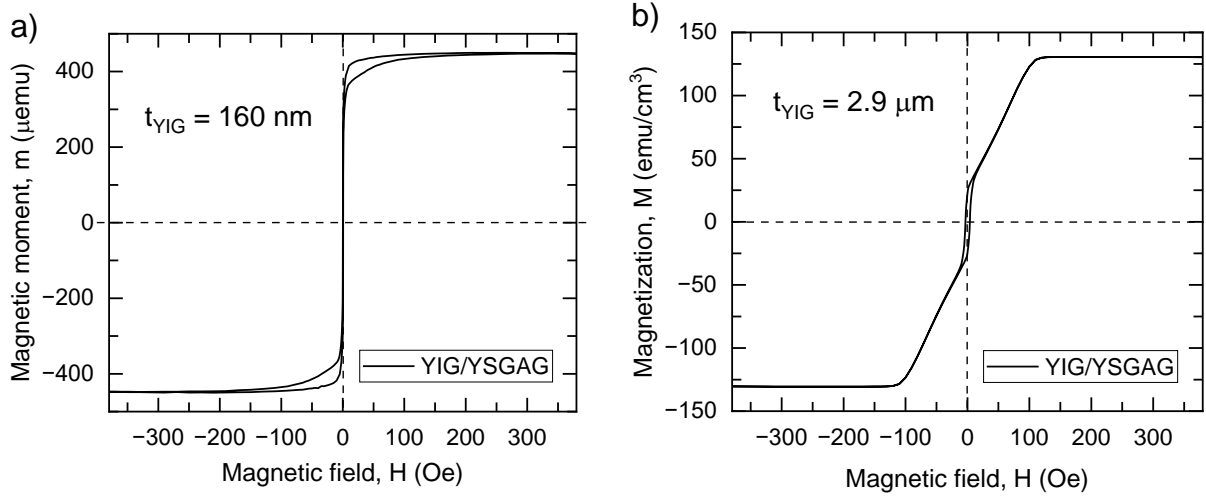

Figure 14: VSM magnetization curves as a function of the in-plane magnetic field $H$, (a) of a 160 nm thin YIG film on a YSGAG substrate with very low remanence $H_c$ and (b) of a 2.9 µm thick YIG film on a YSGAG substrate showing an abrupt increase in magnetization at low bias fields (hysteresis) and a continuous increase up to magnetic saturation of the samples at higher fields.

### 3.10 Microwave characterization

In microwave experiments the FMR linewidth $\Delta H$ is used to identify losses in ferromagnetics. Figure 15 shows examples of frequency spectra and the evaluation of the full width at half maximum (FWHM) of the resonance rf-power absorption peak. Comparatively narrow linewidths of $\Delta f_{FWHM} \sim 9$ MHz can be seen for the YIG/YSGAG sample (see Fig. 15a), while the YIG/GGG sample exhibit $\Delta f_{FWHM} \sim 10$ MHz. The calculated resonance linewidths at 10 GHz and room temperature are $\Delta H_{FWHM} = 3.2$ Oe and = 3.4 Oe, respectively, which is relatively low for inhomogeneous YIG film samples. For more homogeneous films, which are then to be grown on round formatted, perfect epi-polished and defect-free substrates, we expect room temperature values of $\Delta H_{FWHM} \leq 2$ Oe at 10 GHz, which can typically be achieved with the horizontal LPE dipping technique on GGG substrates [27,28]. However, the main problem of YIG films on paramagnetic substrates such as GGG is their damping behavior at cryogenic temperatures, as reported for example in Will-Cole et al. [33] for films grown by different deposition methods. In addition to interfering contributions from paramagnetic impurities incorporated in the film and film defects, the GGG substrates itself induces increasing magnetic stray fields at low temperature [2,3,4]. In order to avoid these stray fields, diamagnetic substrates are essential as mentioned above.

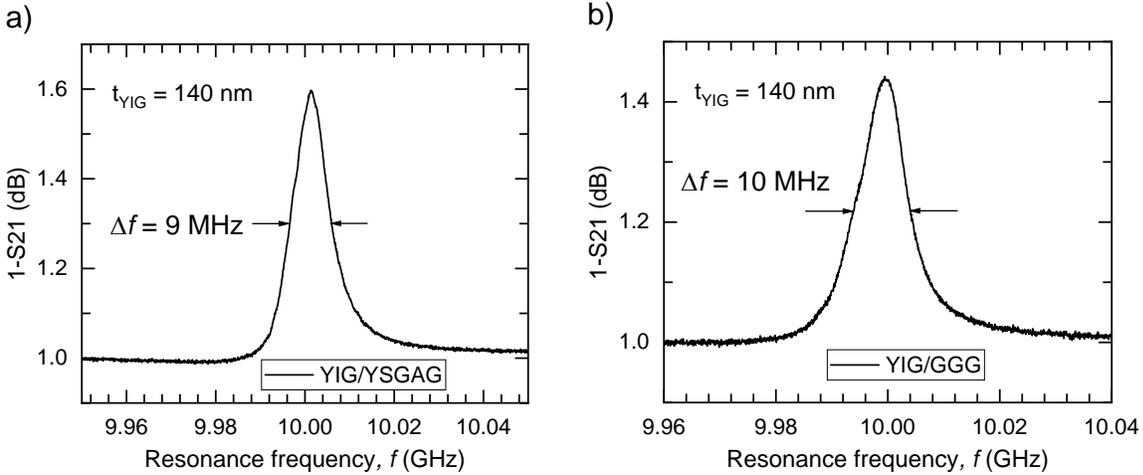

Figure. 15: FMR absorption spectrum close to 10 GHz measured at room temperature using a vector analyzer connected to a broadband stripline a) for a 140 nm YIG film on an YSGAG substrate and b) for a 140 nm YIG film on a GGG substrate.

Since the well-known yttrium aluminum garnet (YAG) has too small lattice parameters and therefore too large stress-induced anisotropy contributions, deposited YIG films onto these substrates show very large FMR linewidths ($\Delta H_{FWHM}$ ~35 Oe @10 GHz) even at room temperature [34]. For this reason, Guo et al. [29] recently demonstrated the growth of YIG epitaxial films by off-axis sputtering on a YSGG single-crystal substrate while avoiding paramagnetic rare-earth elements. For 30 nm thin films, the authors reported FMR linewidths of $\Delta H_{FWHM}$ ~ 27 Oe at 10 GHz and temperatures of 10 K. However, these values are still significantly higher than the values already obtained for LPE YIG films on common paramagnetic GGG substrates at 10 K with linewidths of $\Delta H_{FWHM}$ ~10 Oe for 8 GHz [33]. The same applies to off-stoichiometrically substituted $Y_3(Y_xFe_{5-x})O_{12}$ films [28] at temperatures of 10 K with linewidths of $\Delta H_{FWHM}$ ~ 85 Oe at 10 GHz. The reason for this linewidth broadening seems to be due to rare-earth impurities in the films themselves, as in contrast to the LPE deposition process, where $Y_2O_3$ with a purity of 99.999% was used, it is probable that starting materials of insufficient purity were used for the sputtering target. Another possible reason could be the larger lattice misfit between the YIG film and the YSGG substrate of 0.4% [29] compared to the smaller misfit of 0.2% for the current YIG/YSGAG system. In the case of off-stoichiometrically substituted YIG films, additional defect-related magnetic losses can occur [30].

Initial temperature-dependent FMR experiments for YIG LPE films on YSGAG substrates show that the FMR linewidth is almost not temperature-dependent even at temperatures below 10 K, while the well-known increase in FMR linewidths can be observed for YIG LPE films on common GGG samples grown under the same conditions (see Figure 16). With FMR linewidth values of $\Delta H_{FWHM} \sim 3$ Oe at temperatures below 10 K, they are at least three times lower than for the best YIG LPE films on conventional GGG substrates reported so far [33] and at least one order of magnitude lower than for the best sputtering films on diamagnetic YSGG substrates at 10 K [29]. The observed increase in linewidth in the temperature range between 10 K and 100 K is probably due to the incorporation of paramagnetic rare earth relaxer ions [35] from the initial starting materials and can be further reduced by using more pure rare earth oxides. Therefore, the absolute microwave losses at cryogenic temperatures, especially at millikelvin temperatures, has been significantly reduced by the new YSGAG substrates. Additionally, for more homogeneous and purer YIG LPE films, we expect a further reduction of the microwave losses over the whole temperature range. However, further investigations are required to compare the inhomogeneous linewidth broadening $\Delta H_0$ and the Gilbert damping coefficients $\alpha$, which will be reported in Ref. [36].

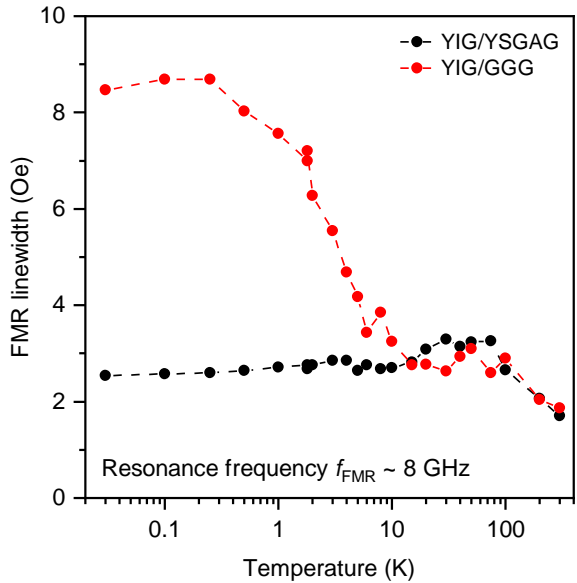

Figure 16: Temperature-dependent FMR measurements for the approximately 140 nm thin LPE films of the YIG/YSGAG and YIG/GGG samples at 8 GHz, deposited under the same conditions using the vertical dipping technique.

4. Conclusions and outlook

First large garnet type $Y_3Sc_2Ga_3O_{12}$-$Y_3Sc_2Al_3O_{12}$ and $Y_3Sc_2Ga_3O_{12}$-$Y_3Al_5O_{12}$ (YSGAG) solid solution single crystals with diameters up to 30 mm and total lengths up to about 100 mm with improved lattice match for YIG films were demonstrated. The novel substrate-film combination led to superior magnetic properties at low temperatures, and due to its potential for a series of applications, a patent application is ongoing (US reference number: 19/295,255). In order to produce larger and core free YSGAG single crystals, future work will be focused on optimizing the crystal growth conditions. To undertake this kind of development, detailed knowledge of the solid-liquid interface shape in dependence of the rotation rate and thermal insulation properties is crucial and therefore currently under investigation.

In addition, the first YIG LPE films with film thicknesses in the submicrometer as well as micrometer range were grown on the first YSGAG substrates, which exhibit significantly reduced microwave damping at low temperatures compared to YIG films on conventional GGG substrates. In order to further improve the homogeneity of the YIG films and their damping properties at low temperatures, double-sided epi-polished substrates with even better adapted lattice parameters and diameters of up to one inch will be used for future deposition series, which can then be processed under substrate rotation by means of the horizontal dipping technique. This should lead to homogeneous and structurally perfect epitaxial films with further reduced FMR linewidths at cryogenic temperatures. These will then form the basis for a novel substrate thin-film system with suitable low-temperature properties on which miniaturized, non-reciprocal microwave elements such as isolators and circulators for the GHz frequency range can be realized.

The material presented here is crucial for the emerging field of quantum magnonics, as it enables operations with long-lived entangled single magnons that propagate in nanoscale solid-state quantum circuits [36].

Acknowledgements

CD would like to thank Th. Friedrich and M. Diegel for the XRD measurements and R. Meyer for the technical support.

The work was supported by the Federal Ministry of Research, Technology and Space (BMFTR) under the reference numbers (13N17108 and 13N17109) within a collaborative project "Low-loss materials for integrated magnonic-superconducting quantum technologies (MagSQuant)". This research was funded in part by the Austrian Science Fund (FWF) project Paramagnonics (10.55776/6568).